\documentclass[prc,twocolumn,showpacs,preprintnumbers]{revtex4}
\usepackage{bm}
\usepackage{amssymb}
\usepackage{amsmath}
\usepackage{graphicx}

\begin{document}
\title{Detecting neutrinos from black hole neutron stars mergers}

\author{O. L. Caballero}\email{olcaball@ncsu.edu}
\author{G. C. McLaughlin}\email{gail_mclaughlin@ncsu.edu} 
\affiliation{Department of Physics,
             North Carolina State University, Raleigh, NC 27695}
\author{R. Surman}\email{surmanr@union.edu}
\affiliation{Department of Physics and Astronomy, Union College, Schenectady, NY 12308}

\date{\today}
\begin{abstract}
While it is well known that neutrinos are emitted from standard core
collapse protoneutron star supernovae, less attention has been focused
on neutrinos from accretion disks.  These disks occur in some supernovae
(i.e. "collapsars") as well as in compact object mergers, and they emit
neutrinos with similar properties to those from protoneutron star
supernovae.  These disks and their neutrinos play an important role in
our understanding of gamma ray bursts as well as the nucleosynthesis
they produce.  We study a disk that forms in the merger of a black hole
and a neutron star and examine the neutrino fluxes, luminosities and
neutrino surfaces for the disk.  We also estimate the number of events
that would be registered in current and proposed supernova neutrino
detectors if such an event were to occur in the Galaxy.
\end{abstract}
\smallskip
\maketitle

\section{Introduction}
Black hole - neutron (BH-NS) star mergers are potential progenitors of short duration gamma ray bursts (GRBs) and have been speculated to be the site of interesting nucleosynthesis. The neutrino emission from the accretion disk produced in BH-NS mergers plays an important role in each of these scenarios. One possible explanation for the energetic GRBs suggests as engine a black hole (BH) of several solar masses accreting matter from a disk. Simulations of compact object mergers have shown the formation of such disks \cite{MacFadyen1999, Taniguchi2005, Lee1999, Rosswog}.  Neutrino transport, annihilation  and losses in the disk material would determine the GRBs production \cite{Matteo02,Popham1999, Setiawan04}.
Several studies have concluded that high accretion rates would provide the necessary conditions for triggering GRBs \cite{Popham1999,SetiawanBHmerger, RuffertGRB-BH}. 
Another interesting aspect is the resulting nuclear products from the accretion disk around black holes (AD-BH). Lattimer and Schramm \cite{Lattimer1974, Lattimer1976} speculated the possibility of a $r$-process in the accretion disk resulting of compact object mergers.  Material from inner crust of the merging neutron star (NS), with low proton fraction,  can be ejected from tidal tails giving place to a least a weak $r$-process.  Furthermore, hot accretion disk winds can also produce an r-process, due to the neutrino interactions \cite{Surmanrprocess}.

Given the conditions of high temperature and density of AD-BH, we expect a copious amount of neutrinos  in the range of 10s of  MeV to be emitted. There are several detectors, both in operation and proposed, that could register 
neutrinos in this energy range, including
 
Super-Kamiokande (SK) \cite{SK}, AMANDA \cite{amanda,icecube}, KamLand \cite{kamland}, ICARUS \cite{icarus}, Ice Cube\cite{amanda,icecube}, LANNDD\cite{lanndd}, and HALO\cite{halo}.  These detectors have been studied extensively for their ability to see
a Milky Way supernova signal.  As detection of the next Galactic core collapse supernova is now within reach experimentally, we can begin to speculate on future detections of even rarer events such as BH-NS mergers and neutron star - neutron star (NS-NS) mergers. Roughly, NS-NS mergers are three orders of magnitude more rare than core collapse supernovae, and BH-NS mergers perhaps another order of magnitude still.  Given the rarity of these events, direct detection on a time scale of years would require a very large detector as we will discuss.  Nevertheless it is in interesting to examine the energy and strength of a signal that would originate from a BH-NS merger and compare with the signal of neutrinos emitted from a proto neutron star (PNS) at the center of  core collapse supernova.

 Previous investigations of MeV scale 
signals in terrestrial detectors of neutrinos which originate from
 black hole accretion disks have focussed on disks that might form from
the core collapse of a rotating massive star.
Nagataki et al, investigated the luminosity, spectrum and counts at SK, for neutrinos from a collapsar \cite{Nagatakicounts}. These authors used an analytical shape for the disk and from it derived the neutrino spectrum. Their results predict at least one neutrino event measured in the proposed 2Mega-ton water Cherenkov  detector, TITAND \cite{Titand}, originating from  an AD-BH at 3 Mpc, when the total accretion mass, the initial mass, and the mass accretion rate are set to $30 M_\odot$, $3M_\odot$, and $0.1M_\odot$s$^{-1}$, respectively. McLaughlin and Surman have considered the possible distinction of neutrino spectra when neutrinos originated in AD-BH versus a PNS \cite{McLaughlin07}.  The AD-BH signal was found to have comparable energy spectra, but the disks primarily emit electron neutrinos and electron antineutrinos, which in addition to the timing of the signal could produce a unique signature after neutrino flavor transformation has been taken into account.

 Determining the neutrino signal from a BH-NS merger is more complex.  In addition to calculating neutrino emission surfaces and estimating the effects of neutrino flavor transformation, general relativistic corrections are more important and one must determine world lines for the neutrinos which originate from different parts of the disk.
In this paper we make estimates of neutrino events registered in several detectors for an AD-BH. We use a hyperaccreting disk model provided by Ruffert and Janka \cite{Surmanrprocess,Ruffert2001}. We also calculate neutrino luminosities and fluxes for the disk. We take into account general relativity corrections for the neutrino energy and disk size. We discuss the consequences of neutrino oscillations in the fluxes and in our event counts. This paper is organized as follows: in section \ref{disk model} we introduce the disk model, in section \ref{neutrino surfaces} we present the reactions included to calculate neutrino surfaces. In section \ref{Flux, Luminosity and Energy} we discuss our results for neutrino fluxes, energies and luminosities. In section \ref{Neutrino Counts} we consider events rates in different detectors, in section \ref{Neutrino Mixing} we take into account neutrino mixing, and in section \ref{Conclusions} we discuss our conclusions.

\section{Disk Model}
\label{disk model}

The results of our calculations are mainly based on a 3D hydrodynamic simulation. However, for comparison purposes, we also investigate results based on a steady state disk. Both models are briefly  described below.

For the hydrodynamical model, we use the simulation results of a BH-NS merger by Ruffert and Janka  \cite{ Setiawan04, Eberl99, Surmanrprocess,Ruffert2001}, for a $1.6 M_{\odot}$ NS and a $2.5 M_{\odot}$ BH with spin parameter $a=Jc/GM^2=0.6$ ($J$ is the total angular momentum and $M$ the rest mass of the system). These authors follow the hydrodynamics of the merger with the Piecewise Parabolic Method \cite{Ruffert1996}, including gravitational wave emission and neutrino emission \cite{Ruffert2001}.

 In this model, general relativistic effects are included by using a modified Newtonian potential. The BH is treated as a gravitational centre surrounded by a vacuum sphere. The gravitational potential $\Phi_{BH}$ of the BH is an extension of the Paczynski-Wiita potential \cite{BHpote} to rotating BH \cite{Artemova}.  As function of radius $r$, $\Phi_{BH}$ has the form
 
 \begin{equation}
\frac{d\Phi_{BH}}{dr}=\frac{GM_{BH}}{r^{2\beta}(r-r_H)^\beta},
\end{equation}

 where $\beta$ depends on the BH spin parameter $a$, and $r_H$, $M_{BH}$ are the event horizon and mass of the BH respectively. Artemova et al. \cite{Artemova} compare the results obtained using this potential to the results obtained with a more complete genereal relativistic treatment and find similar disk structures at 10\%-20\% level. 
 
 In the model of Ruffert and Janka the Shen et al.\cite{ShenEoS} equation of state is used to described the stellar matter. The simulation was evolved until the accreted material formed a disk around the BH. Therefore,  our analysis is based on one snapshot of the disk's evolution.
The starting points for our calculation are the temperature $T$, density $\rho_m$ and electron fraction $Y_e$ results for each of the coordinates $\rho$, $\phi$ and $z$ of a cylindrical grid. The inner boundary of the model is located at $\rho=14$ km and the disk surface extends until $\rho=300$ km.  While neutrino emission is already included in the numerical model, for our purposes we would like more detailed information about the neutrinos, so in the following section we describe a ``post-processing'' of the output from this model.

To compare with a steady state disk we use the model of Chen and Belobodorov \cite{Beloborodovcross}. This model is fully relativistic. The disk is one dimensional, axially symmetric and is described by vertically averaged quantities. These authors worked with two different values of the spin parameter $a=0$ and $a=0.95$, which we also use here. The mass of the BH is 3$M_\odot$ and the accretion rate $M=5M_\odot/$s. The disk extension goes as far as $\rho=600$km.
For the vertical structure of the disk we use a simple hydrostatic model that assumes an equilibrium with the gas radiation pressure and gravity.

\section{Neutrino Surfaces}
\label{neutrino surfaces}

Analogously to the neutrino spheres in a PNS, we can define the surface at which neutrinos decouple from the accretion disk. We follow a procedure similar to that outlined by Surman and McLaughlin in Ref \cite{Surman_Mclaughlin04}. The disk can be divided in regions according to their neutrino opacity. If in a given region the optical depth $\tau_\nu> 2/3$, then neutrinos are trapped and the disk is said to be optically thick. In the region where $\tau_\nu < 2/3$ the disk is optically thin to neutrinos. Unlike in a PNS, neutrino surfaces in a disk depend on the direction of neutrino emission. We are interested in calculating neutrino surfaces directly above the equatorial plane of the disk.  
Therefore, we find the height of the neutrino surface,  $h_{\nu}$ at each $\rho$, $\phi$
by changing the lower limit in the integral
\begin{equation}
\label{opticaldepth}
\tau_{\nu}=\int^{h_{max}}_{h_{\nu}}\frac{1}{l_{\nu}(z)}dz,
\end{equation}
such that $\tau_\nu=2/3$. Here $h_\nu$ corresponds to a $z$ value in the cylindrical grid, $h_{max}$ is the maximum distance in the $z$ direction where matter is found and $l_\nu$ is the neutrino mean free path which is given by 
\begin{equation}
\label{meanfreepath}
l_\nu=\frac{1}{\sum_kn_k\langle\sigma_k(E_\nu)\rangle}.
\end{equation}
The summation in the above equation runs over different neutrino scattering process, which we describe below, $n_k=n_k(\rho,\phi,z)$ is the associated particle density  of each process  and 
\begin{equation}
\label{averagedcross}
\langle  \sigma_k(E_\nu)\rangle=\frac{\int ^\infty_0\sigma_k(E_\nu)\phi (E_\nu)dE_\nu} {\int^\infty_0\phi(E_\nu)dE_\nu}
\end{equation}
is the  corresponding cross section averaged over the Fermi-Dirac flux
\begin{equation}
\phi(E_\nu)=\frac{g_\nu c}{2\pi^2 (\hbar c)^3}\frac{E_\nu^2}{\exp(E_\nu/T)+1},
\label{flux}
\end{equation}
with $g_\nu=1$ and with an assumed neutrino chemical potential $\mu_\nu=0$. Strictly speaking the flux $\phi(E_\nu)$ is also a function of $T$ and therefore a function of $\rho$ as well as of $\phi$ and $z$. We will point out this dependency when needed by noting the value of $T$ at which $\phi$ is evaluated.

As matter is dragged into the black hole the medium becomes hotter and denser, and nuclei dissociate.  Therefore, we consider neutrino scattering from protons, neutrons and electrons. We have the current charged reactions for electron (anti)neutrino $(\bar{\nu_e})\nu_e$:
\begin{equation}
\nu_e +n\rightarrow p+e^-
\label{nuabsorption}
\end{equation}
\begin{equation}
\label{anuabsorption}
\bar{\nu}_e +p \rightarrow e^+ + n,
\end{equation}
and for all (anti)neutrino flavors the neutral current processes
\begin{equation}
\label{nuproton}
\nu+p \rightarrow \nu+p,
\end{equation}
\begin{equation}
\label{nuneutron}
\nu+n \rightarrow \nu+n,
\end{equation}
\begin{equation}
\label{nuelectron}
\nu+e^- \rightarrow \nu+e^-,
\end{equation}
\begin{equation}
\label{annihilation}
\nu+\bar{\nu} \rightarrow e^++e^-.
\end{equation}
The cross section for neutrino absorption (Eq, \ref{nuabsorption}), including weak magnetism effects $W_M$ \cite{Horowitzwm} is
\begin{eqnarray}
\label{absorption}
\sigma_{\nu_en\rightarrow pe^-} &=& \frac{\sigma_0}{4m^2_e}(1+3g^2_A)(E_\nu+\Delta)^2\nonumber\\
&\times&\left[1-\left(\frac{m_e}{E_\nu+\Delta}\right)^2\right]^{1/2}W_M,
\end{eqnarray}
where
\begin{equation}
W_M=\left(1+1.1\frac{E_\nu}{m_n}\right),
\end{equation}
$g_A=1.93$, $m_{e(n)}$ is the electron(neutron) mass, $\Delta=1.23$ is the neutron proton mass difference, and $\sigma_0=4G^2_Fm^2_e/\pi \hbar^4$. 
For scattering of electron antineutrinos from protons, Eq. \ref{anuabsorption} we have
\begin{eqnarray}
\label{absorption}
\sigma_{{\bar\nu}_ep\rightarrow ne^+} &=& \frac{\sigma_0}{4m^2_e}(1+3g^2_A)(E_\nu-\Delta)^2\nonumber\\
&\times&\left[1-\left(\frac{m_e}{E_\nu-\Delta}\right)^2\right]^{1/2}W_M,
\end{eqnarray}
with
\begin{equation}
W_M=\left(1-7.1\frac{E_\nu}{m_n}\right).
\end{equation}

The cross section for the neutral current process of Eq. \ref{nuproton}, valid for all neutrino flavors is \cite{Burrows}
\begin{equation}
\sigma_{\nu p\rightarrow\nu p}=\frac{\sigma_0[(C_V-1)^2+3g^2_A(C_A-1)^2]}{4m^2_e}E_\nu^2,
\end{equation}
with $C_V=1/2+2\sin^2\Theta_W$, the Weinberg angle $\sin\Theta^2_W=0.23$ and $C_A=1/2$, whereas the corresponding expression for Eq.\ref{nuneutron} is
\begin{equation}
\sigma_{\nu n\rightarrow \nu n}=\frac{\sigma_0(1+3g^2_A)}{16m^2_e}E^2_\nu.
\end{equation}
The same expressions hold for antineutrinos if we change $g_A=-g_A$. This change does not have consequences in the above expressions.

In order to obtain $\sigma(E_\nu)$ for the electron neutrino elastic scattering from electrons (Eq. \ref{nuelectron}), we integrate the differential cross section \cite{Bahcall}
\begin{equation}
\begin{split}
\frac{d\sigma_{\nu e}}{dT_e}=\frac{\sigma_0}{8m_e}\left[(C_V +C_A)^2+(C_V-C_A)^2\left(1-\frac{T_e}{E_\nu}\right)^2\right. \\
\left.-(C^2_V-C^2_A)\frac{m_eT_e}{E^2_\nu}\right],
\end{split}
\end{equation}

where,
\begin{equation}
T_e=\frac{(1-\cos\theta)E^2}{m_e+(1-\cos\theta)E}.
\end{equation}
and evaluate it between the maximum and minimum values of $T_e$, corresponding to $T_e=\frac{2E^2}{m_e+2E}$ and 0, respectively. For $\nu_x$ ($x=\mu, \tau$) scattering from electrons we have the same expression with $C_A=-1/2$ and $C_V=2\sin^2\Theta_W -1/2$.  Antineutrino scattering requires changing $C_A=-C_A$.

Finally, an approximate form for the cross section for neutrino-neutrino annihilation, Eq.\ref{annihilation}, is taken from Dicus \cite{Dicus} and Goodman \cite{Goodman87}, assuming $E_\nu, E_{\bar\nu} >> m_e$ \cite{Beloborodovcross} (for the complete expression see  \cite{Kneller:2004jr}), 
\begin{equation}
\sigma_{\nu{\bar\nu}\rightarrow ee^+}(E_\nu)=\frac{4}{3}K_{\nu{\bar\nu}}\sigma_0E_\nu\langle E_{\bar\nu}\rangle,
\end{equation}
for neutrinos and
\begin{equation}
\sigma_{\nu\bar{\nu} \rightarrow ee^+}(E_{\bar\nu})=\frac{4}{3}K_{\nu\bar{\nu}}\sigma_0E_{\bar{\nu}}\langle E_\nu\rangle,
\end{equation}
for antineutrinos. Here $\langle E_{\nu({\bar\nu})}\rangle$ are the average (anti)neutrino energies and
\begin{equation}
K_{\nu_e\bar\nu_e}=\frac{1+4\sin^2\theta_W+8\sin^4\Theta_W}{6\pi},
\end{equation}

\begin{equation}
K_{\nu_x\bar\nu_x}=\frac{1-4\sin^2\theta_W+8\sin^4\Theta_W}{6\pi}.
\end{equation}

To calculate the proton $n_p$ and neutron $n_n$ number densities needed in Eq. \ref{meanfreepath}, we start from the electron fraction fraction $Y_e$ and the mass density $\rho_m$  of the disk model. We assume charge neutrality $Y_e=Y_p$ so we have $n_p=\rho_m N_AY_e$ and $n_n=\rho_m N_A(1-Y_e)$, where $N_A$ is the Avogadro's number. We find the electron number density $n_e$ using
\begin{equation}
\mu_{e^-}+\mu_{e^+}=0,
\end{equation}
and Fermi-Dirac distributions for electrons and positrons with the corresponding grid values of temperature. We proceed by finding $\mu_{e^-}$ such that $n_p=n_{e^-}-n_{e^+}$, and with this value of $\mu_{e^-}$ we find from the Fermi-Dirac distribution that characterizes $n_e$ for each grid point.

\begin{figure}[ht]
\begin{center}
\includegraphics[width=3.75in,angle=0,clip=true] {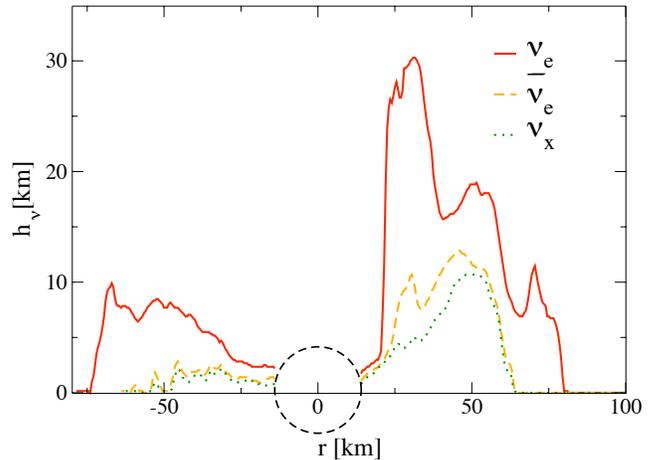}
\caption{(Color on line) Transversal cut of neutrino surfaces at $\phi=20^\circ$. The solid line corresponds to electron neutrino surface whereas the dashed and dotted lines correspond to electron antineutrino and tau neutrino respectively. The circle around $r=0$ represents the BH boundary.}
\label{nusurfacecut}
\end{center}
\end{figure}

 The procedure described above allows us to find the temperatures $T_\nu$ at which neutrino decouple. Those correspond to the temperatures at $h_\nu$ in Eq. \ref{opticaldepth}. We show the result of this calculation in Fig. \ref{nusurfacecut}. This figure shows a  transversal cut of the resulting neutrino surface corresponding to a polar angle $\phi=20^\circ$ in the original numerical simulation. 

The first fact to be noticed is its irregular shape, in contrast to the symmetric one of a neutrino sphere.  
 Also, because the material in the disk is relatively neutron rich, $\nu_e$ absorption on neutrons has a more significant contribution to the mean free path, than $\bar{\nu}_e$ absoprtion on protons.  Thus the electron neutrinos decouple at the lowest temperatures. The mu and tau type neutrinos and antineutrinos lack these charged current interactions and decouple at the highest temperatures. This is the same hierarchy of energies that is seen in neutrinos emitted from the PNS.

\begin{figure}[ht]
\begin{center}
\includegraphics[width=3.3in,angle=0,clip=true] {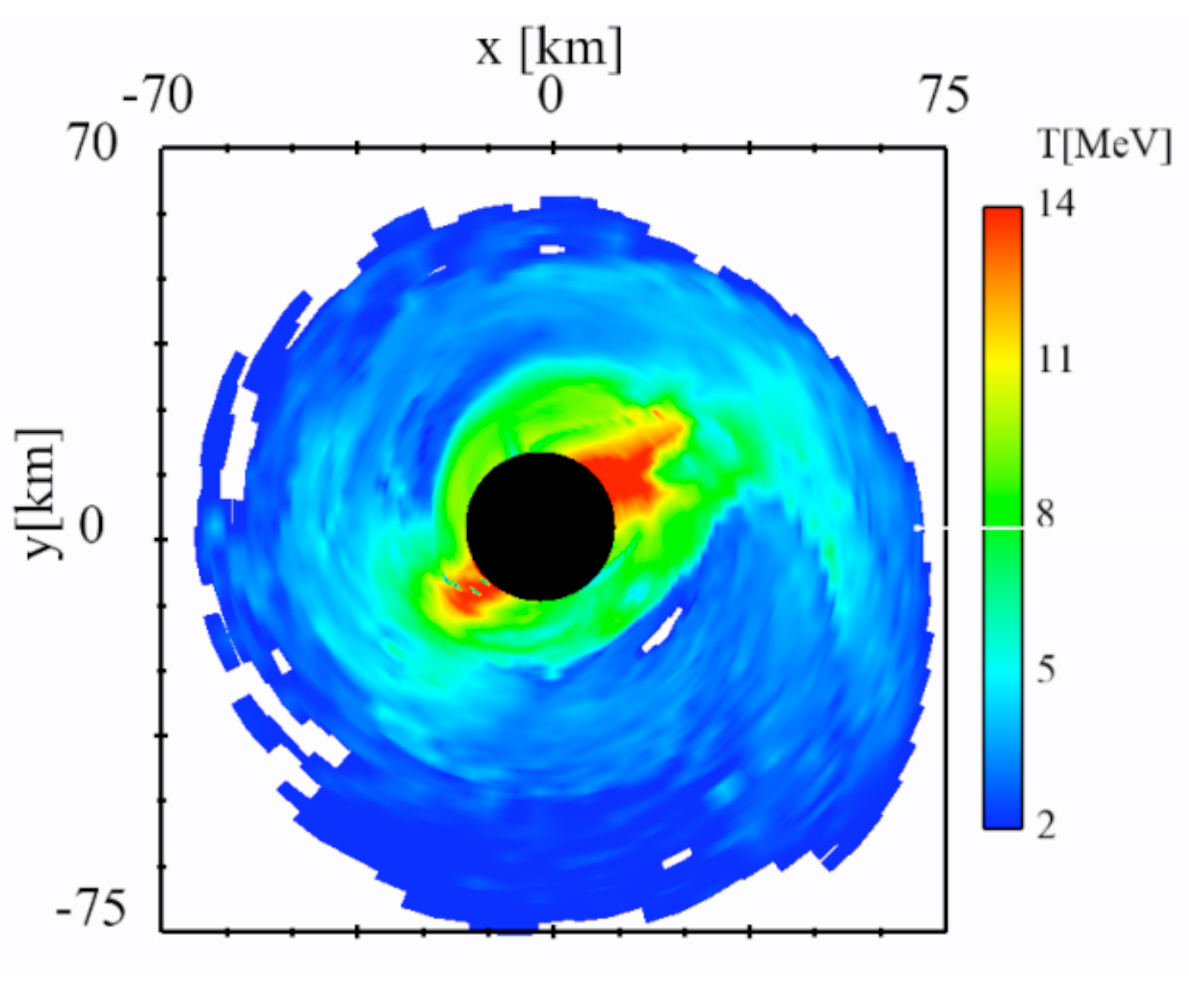}
\caption{(Color on line) Temperature profile of the electron antineutrino surface seen from the $z$ axis. The black frame, $x=[-70, 75]$, $y=[-75,70]$ km, encloses the antineutrino surface. The temperature scale (on the left) goes from blue $T=2$ MeV, to red $T\sim14$ MeV. The black circular area represents the black hole boundary, $r=2r_s$.}
\label{nuT}
\end{center}
\end{figure}

 Fig. \ref{nuT} shows the electron antineutrino surface temperatures for the whole disk. The disk is in the $x,y$ plane and the color scale represent the temperatures. Blue corresponds to $T=0$. The hotter ${\bar\nu_e}$ are closer to the BH (black circle in the center). These temperatures are higher than the PNS. While the core temperatures in a PNS are of the order of hundred MeV, the temperatures of the AD could be around 20 MeV. However, the density profile of the disk is very different from that of a PNS. When a neutrino decouples in a PNS it has diffused through a denser medium which is assumed to be more or less symmetric in all directions, giving as a result a spherical shell for the neutrino surface with lower temperature. In the case of the disk, as matter is dragged to the BH, a funnel is formed around the BH vicinity. Therefore, changes in density are more abrupt. The medium is less dense close the BH, increases rapidly between 30 to 60 km and then decreases again as $r$ increases. 
Neutrinos emitted close to the BH travel through a less dense medium for less time, resulting in higher temperatures compared to those in a PNS. 

\begin{figure}[ht]
\begin{center}
\includegraphics[width=3.3in,angle=0,clip=true] {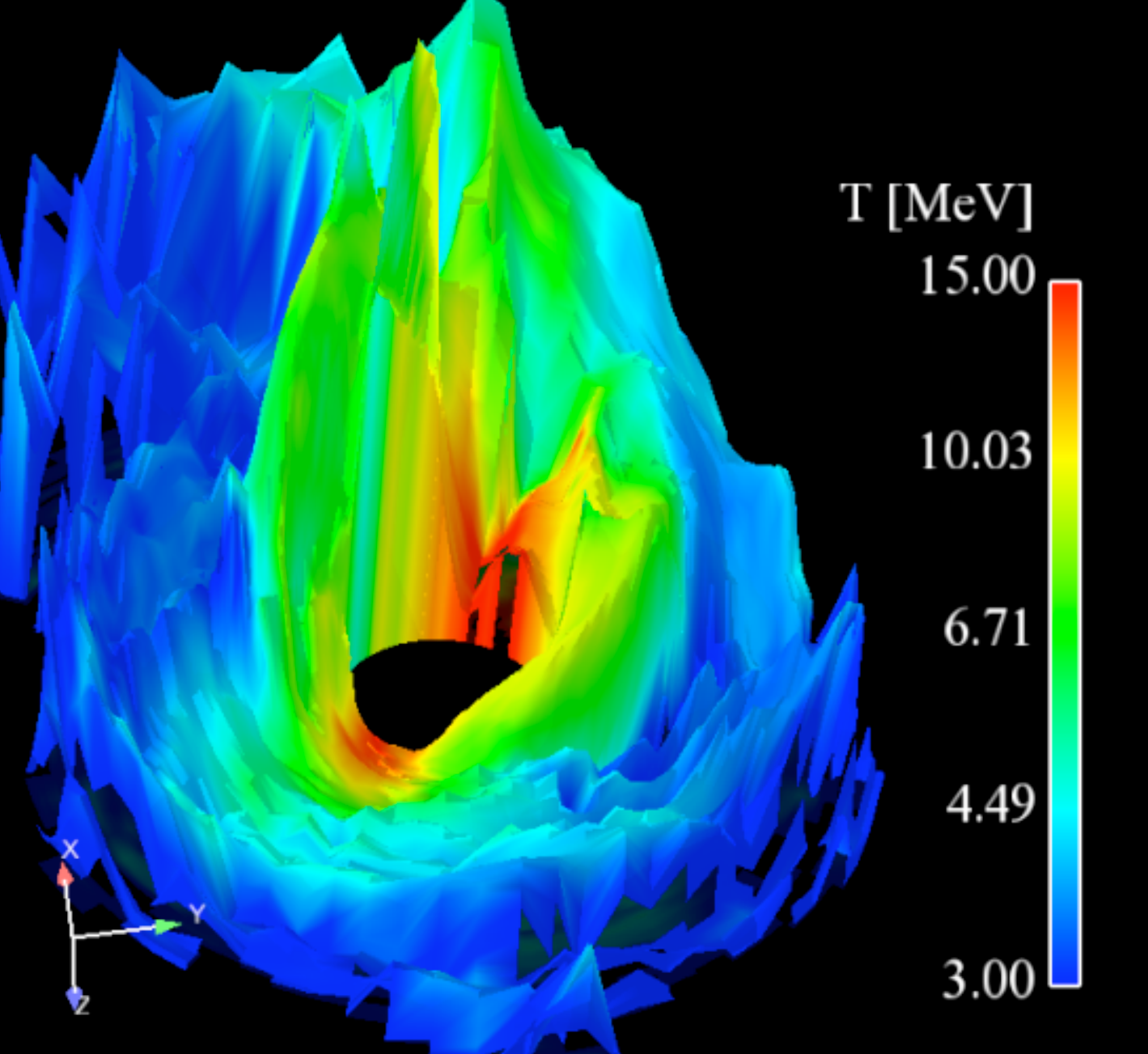}
\caption{(Color on line) Electron antineutrino surface seen at some inclination angle (see the $x$, $y$, $z$ axis on the lower left  corner). The height corresponds to $h_\nu$ as in Eq. \ref{opticaldepth}. The color scale corresponds to the neutrino temperatures, also shown in Fig. \ref{nuT}. The black area in the center represents the boundary with the BH, $r=2r_s$.}
\label{nuzT}
\end{center}
\end{figure}

 Fig. \ref{nuzT} shows a 3D image of the $\bar{\nu}_e$ surface. The height represents $h_\nu$ while the color scale represents $T_{{\bar\nu}_e}$. The biggest contribution to the antineutrino flux comes from regions closer to the boundary with the BH. There, the $h_\nu$ is smaller which translates in higher decoupling temperatures. 
 As described in the text, this is a consequence of the density profile. Regions where $h_\nu$ is high and far away from the center contribute less to the flux.   
 
\section{Flux, Luminosity and Energy}
\label{Flux, Luminosity and Energy}
Using the values of $T_\nu$ we calculate fluxes, luminosities and averaged energies for each neutrino flavor. For the neutrino luminosity  we integrate $E_\nu$ over the neutrino surface, assuming the Fermi-Dirac distribution  of Eq. \ref{flux},
\begin{equation}
\label{luminos}
L_\nu=\int_0^{2\pi}d\phi\int_{\rho_{min}}^{\rho_{max}}\rho d\rho\int^\infty_0E_\nu \phi(E_\nu,T_\nu)dE_\nu.
\end{equation}
Here $\rho_{min}=14$ km, $\rho_{max} $ corresponds to the boundary, in the radial direction, on the $x,y$ plane, of the optically thick region, and $\phi$ is the polar angle. $L_\nu$ is the total energy emission rate and does not depend on a specific observation point.  The total luminosity  is obtained by adding the resulting luminosity for each neutrino flavor. Similarly, to estimate the total number of neutrinos emitted per unit time $f$,  we integrate the flux over the neutrino surface,

\begin{equation}
f=\int_0^{2\pi}d\phi\int_{\rho_{min}}^{\rho_{max}}\rho d\rho\int_0^\infty\phi(E_\nu,T_\nu)dE_\nu.
\end{equation}

Our estimate for the average neutrino energy is then given by
\begin{equation}
\langle E_\nu\rangle=\frac{L_\nu}{f}.
\end{equation}

Because of the presence of the black hole the above quantities will differ from those measured by an observer at infinity.  A source located at a distance $r$ from a black hole emitting at energy $E$ will be observed at infinity to have an energy, \cite{gravitation},
\begin{equation}
E^*= \frac{E}{1+z}.
\label{eshift}
\end{equation}
The redshift factor $1+z$ consists of a Doppler part and a gravitational part (see Ref. \cite{Luminet} for a derivation). The Doppler term depends on the ratio of $\Omega$, the angular velocity of the emitting gas, to the speed of light. 
We find that the Doppler term is several orders of magnitude smaller than the gravitational term. Therefore, we use
\begin{equation}
\frac{1}{1+z}=|g_{00}|^{1/2},
\end{equation}
and then the energy observed at infinity is
\begin{equation}
E^*= |g_{00}|^{1/2}E,
\label{shiftenergy}
\end{equation}
where $g_{00}$ is determined by the space time metric. In the case of a non-charged, rotating black hole the curvature of the line element can be written in the Kerr geometry as \cite{gravitation},
\begin{eqnarray}
ds^2&=&-(\Delta/\xi^2)\left[dt-a^2\sin^2\theta d\phi\right]^2 \nonumber\\
&&+\left(\sin^2\theta/\xi^2\right)\left[(r^2+a^2)d\phi-adt\right]^2\nonumber\\
&&+\left(\xi^2/\Delta\right)dr^2+\xi^2d\theta^2,
\end{eqnarray}
where
\begin{eqnarray}
\Delta=r^2-r_sr+a^2\nonumber\\
\xi^2=r^2+a^2\cos^2\theta,
\end{eqnarray}
and the Schwarzchild radius $r_s=2M$. $g_{00}$ is given by 

\begin{equation}
g_{00}=-\left(\frac{\Delta-a^2\sin^2\theta}{\xi^2}\right)=-\left(1-\frac{r_sr}{\xi^2}\right).
\label{redshift}
\end{equation}
Note that the parameter $a$ given above is the spin of the black hole, not the disk.  Before the merger the spin parameter of the  black hole is $a=0.6$.
The Kerr metric reduces to the Schwarzschild case when $a=0$. We will use the Kerr metric to describe the  neutrino energy redshift. However, for simplicity (and because the spin parameter introduces only a small correction unless it is nearly one) we use the Schwarzschild metric to calculate corrections due to the neutrino ray bending.  An observer at infinity, from a BH will detect a luminosity $L^*=|g_{00}|L$ from an object which has a luminosity $L$ \cite{Shapiro}. For our disk we calculate the redshift in energy $|g_{00}(r_\nu)|$ as in Eq.\ref{redshift} at the point of decoupling $r^2_\nu=h^2_\nu+\rho^2_\nu$, with $h_\nu$ and $\rho_\nu$ the corresponding cylindrical emission coordinates. Here again $L^*_\nu$  is the total emission rate and does not take into account a specific location of the observer \cite{Birk07:annhihilation}.Then we have,

\begin{equation}
L^*_\nu=\int_0^{2\pi}d\phi\int^{\rho_{max}}_{\rho_{min}} g_{00}(r_\nu)\rho d\rho\int^\infty_0E_\nu \phi(E_\nu,T_\nu)dE_\nu,
\end{equation}
\begin{equation}
\label{obsTflux}
f^*= \int_0^{2\pi}d\phi\int_{\rho_{min}}^{\rho_{max}} g^{1/2}_{00}(r_\nu)\rho  d\rho\int^\infty_0\phi(E_\nu,T_\nu)dE_\nu,
\end{equation}
and
\begin{equation}
\langle E_\nu\rangle^*=\frac{L_\nu^*}{f^*},
\end{equation}

for the observed flux, luminosity and averaged energy respectively.
Table \ref{energytable} shows our results for averaged energies and luminosities, both emitted and observed. For comparison we have added the values corresponding to a PNS. The total observed luminosity for this AD-BH is $1.6\times10^{54}$ ergs/s, a hundred times larger than the luminosity of a PNS, $L\sim10^{52}$ ergs/s.

We find the results for the Chen and Beloborodov steady state disk, applying the same technique for the calculation of neutrino surfaces, energies and luminosities. The energies obtained for spin parameters, $a=0$  and $a=0.95$, are lower than the results of Table \ref{energytable}. For example, the average energy for electron antineutrino is $E_\nu$= 11 MeV and $E_\nu=14.4$ MeV, for $a=0$ and $a=0.95$ respectively. This difference can be understood in terms of the temperature and density dependence with the distance to the BH. For both models the temperatures and densities are  similar in the region close to the BH, where most neutrinos are emitted. The Janka and Ruffert model predicts highly fluctuating temperatures and densities that drop faster with distance. The Chen and Belobodorov model leads to symmetric neutrino surfaces and a smooth temperature decay. Therefore, as the distance increases neutrinos with low temperatures constitute a significant fraction of the final spectra lowering the average energies.
\begin{table}
\begin{center}
\caption{Observed $E^*(L^*)$ and emitted $E(L)$ averaged neutrino energies (luminosities) for an AD-BH and for a PNS.}
\begin{tabular}{llllllll}
\hline
  &$E$(MeV)&$E^*$(MeV) & $E $(MeV)&$L$(ergs/s) &$L^*$ (ergs/s) \\
  &Disk&Disk&PNS\cite{Dighe08}&Disk($\times10^{53}$)&Disk($\times10^{53}$)\\
  \hline
$ \bar{\nu}_e $&29.6&  23.4 & 15 &3.7& 2.4\\
 $\nu_e$ &  21.1 &17.3&  $12$   & 2.3&1.6\\
$ \bar{\nu}_x$&33& 26&$25$ & 4.6&3.0\\
 $\nu_x$  &33&26& $25$ & $4.6$&3.0 \\
\hline
\end{tabular} 
\label{energytable}
\end{center}
\end{table}

\section{Neutrino Spectra}
\label{Neutrino Counts}
So far we have determined neutrino energies and luminosities, both emitted and observed at infinity. In this section we obtain the neutrino fluxes as seen at a fixed point and apply our results to estimate the number of neutrinos registered at that specific point.

The number of neutrinos emitted per unit energy, per unit area, per second, reaching an observer located above the disk plane, on the $z$ axis, at a distance $z_{eva}$ is given by 
\begin{equation}
\phi^{eff}(E^*_\nu)=\frac{1}{4\pi}\int_0^{2\pi}\int^{\theta_{max}}_0\sin\theta d\theta d\phi\times\phi(E^*_{\nu}).
\end{equation}
Here $\theta_{max}$ is the maximum angle formed by the outer edge of the neutrino surface and the $z$ axis, as seen by an observer at $z_{eva}$. $E^*$ is the redshifted energy.

\begin{figure}[ht]
\begin{center}
\includegraphics[width=3.3in,angle=0,clip=true] {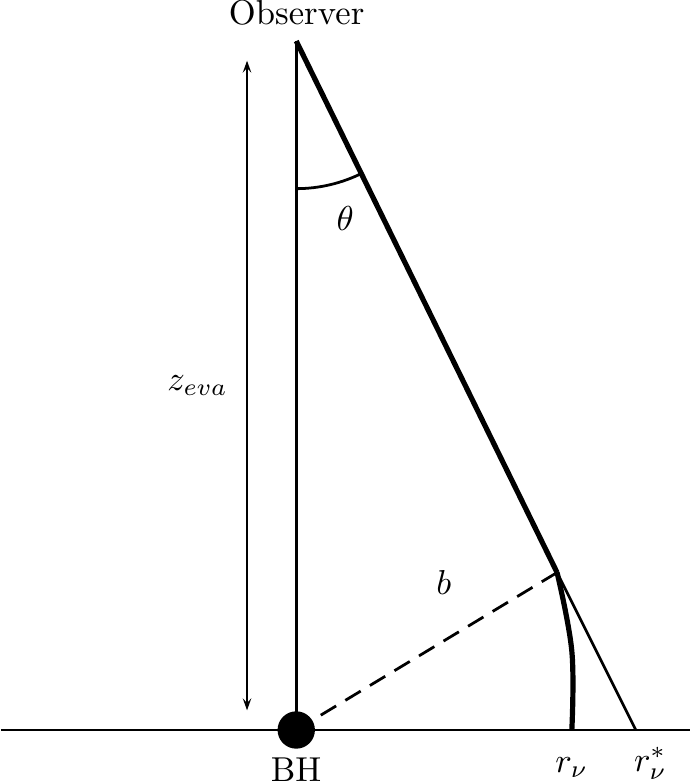}
\caption{Diagram showing the effects of the BH gravitational field on neutrino trajectories. The black dot represents the BH, $r_\nu$ is the emission point on the neutrino surface, which we have sketched with the horizontal line, $b$ is the impact parameter of the trajectory (thick line) and $\theta$ is the angle measured by an observer at $z_{eva}$. According to this observer the emission point would be located at $r^*_\nu$.}
\end{center}
\label{bendray}
\end{figure}

 If the BH did not affect the neutrino trajectories the angle $\theta_{max}$ subtended by an observer at $z_{eva}$ would be given by $\tan(\theta_{max})=r_{max}/z_{eva}$, with $r_{max}$ defined by the boundary of the neutrino surface. However, the presence of the BH does bend the trajectories. To take this into account, we can follow backwards the neutrino trajectories by tracing null geodesics leaving the observer at $z_{eva}$ and reaching the disk\cite{T.Muller}. The angle subtended by the disk according to the observer can be calculated in terms of the impact parameter $b$, which is a constant over the trajectory. The diagram in Fig. \ref{bendray} shows the effect of the BH on the neutrino trajectories.  Neutrinos leave the neutrino surface at the emission point $r_\nu=(h^2_\nu+\rho^2_\nu)^{1/2}$. Their trajectories bend according to their separation from the BH. The influence of the gravitational field is less strong when the neutrinos are far from the BH. $b$ can be visualized by assuming that when the neutrino is far away from the BH it travels in a straight line. The impact parameter is the distance between the closest approach of the continuation of this straight line and the center of the BH \cite{Nemiroff}. 

According to the definition of $b$ we have,
\begin{equation}
\sin\theta=\frac{b}{z_{eva}},
\end{equation}
and therefore for the element of solid angle seen by an observer at infinity $d\Omega=bdbd\phi/z^2_{eva}$ \cite{Bhattacharyya}. On the other hand, $b$ that a neutrino going straight up will encounter at the emission point $r_\nu$ is in the Schwarzschild metric given by
\begin{equation}
b=\frac{r_\nu}{(1-r_s/r_\nu)^{1/2}},
\end{equation}
and
\begin{equation}
db=\frac{dr_\nu}{(1-r_s/r_\nu)^{1/2}}\left[1-\frac{r_s}{2r(1-r_s/r_\nu)}\right].
\end{equation}
Using this expression, the integral over $b$  can be written in terms of  $r_\nu$ (or equivalent $\rho_\nu$) and then the effective flux will be written as
\begin{equation}
\phi^{eff}(E^*_\nu)=\frac{1}{4\pi}\int_0^{2\pi}\int \frac{bdb}{z^2_{eva}} d\phi\times\phi(E^*_{\nu}).
\label{fluxeff}
\end{equation}
with limits given by boundaries of the neutrino surface.

\subsection{Neutrino Counts}

We now use the neutrino fluxes calculated as above to estimate the number of counts in current and proposed neutrino detectors. We assume that we observe the disk from the $z$ axis at distance $z_{eva}=500$ km and then we re-scale our results to 10kpc. We evaluate the count rate $R$, in a given detector, by integrating the effective flux over the energies recorded at the detector
\begin{equation}
R=N_T\int_{E_{th}}^{\infty} \phi^{eff}(E_{\nu}/g_{00}(r))\sigma(E^*)dE^*.
\end{equation}

Here $N_T$ is the number of targets in the detector and $E^*$ and $E$ are related by Eq. \ref{eshift}. The flux $\phi^{eff} (E_\nu/g_{00}(r))$ is the redshifted neutrino distribution of Eq.\ref{fluxeff}, with $\phi(E_\nu)$ evaluated at $T_\nu$,  $E^*$ is the detected energy,  and $E_{th}$ and $\sigma(E^*)$ correspond to the threshold energy and cross section of the detector. Using Eq. \ref{fluxeff} we write $R$ as
\begin{equation}
\label{ratecount}
R=\frac{1}{4\pi}\int_0^{2\pi} d\phi\int \frac{b db}{z_{eva}^2}\int_{E_{th}}^{\infty} \phi(E_{\nu}/g_{00}(r))\sigma(E^*)dE^*.
\end{equation}
The cross sections in Eq. \ref{ratecount} depend on the detector under consideration. For SK this corresponds to the expression Eq.\ref{absorption}. The energy threshold in SK is 5 Mev. Our results are for a 32 kton volume. 

We can easily re-scale our results to the bigger volumes of the proposed detectors UNO (580 kton) \cite{Goodmanuno,uno} and Hyper-K (1 Mton)\cite{Nakamura}. On the other hand, the large scale AMANDA (in operation) and Ice-Cube (in construction) detectors, though designed to detect high energy neutrinos, have been discussed to detect supernova neutrinos if the number of them is large enough to allow extracting their signal from the detector background \cite{amanda,icecube}. We think this is the case for the AD-BH neutrinos and we estimate counts in these detectors using $E_{th}=0$ and an effective volume of $414$ m$^3$ for each of    its 680 optical modules (OM). Ice-Cube is an extension of AMANDA with 4800 OM. 

For supernova neutrino detection, Argon based facilities such as ICARUS or LANNDD  rely on the reactions:
\begin{equation}
\nu_e+^{40}Ar \rightarrow e^-+^{40}K^*,
\end{equation}
and
\begin{equation}
\bar{\nu}_e+^{40}Ar \rightarrow e^++^{40}Cl^*,
\end{equation}
as well as elastic scattering from electrons. We estimate our counts in these detectors assuming a fiducial volume of 3 kt and 70 kt for ICARUS and LANNDD, respectively.  We take the cross sections from Refs. \cite{arcross,KolbeAr}. 

In a lead based detector, such as the proposed HALO, the high number of neutrons Pauli-block the electron antineutrino absorption. However, the neutrino scattering cross section per nucleon is large and most of the events produce neutrons. Neutral current process also occur with the emission of neutrons. We have used the cross sections by  J. Engel et al. \cite{lead} to calculate rate counts for the charged current reactions:
\begin{equation}
\nu_e+^{208}Pb \rightarrow ^{207}Bi+n+e^-,
\end{equation}
\begin{equation}
\nu_e+^{208}Pb \rightarrow ^{206}Bi+2n+e^-,
\end{equation}
and for the neutral current processes,
\begin{equation}
\nu+^{208}Pb \rightarrow ^{206}Pb+2n,
\end{equation}
and
\begin{equation}
\nu+^{208}Pb \rightarrow ^{207}Pb+n.
\end{equation}
KamLAND also offers the possibility of neutrino detection from supernovae via  elastic scattering from protons. Beacom et al. \cite{kamland} predict neutrino proton recoil spectrum for those processes including quenching of the proton scintillation light and different detector backgrounds. We follow their treatment for the proton energy threshold. We start from the differential neutrino-proton cross section
 \begin{equation}
\begin{split}
\frac{d\sigma_{\nu p}}{dT_p}=\frac{\sigma_0 m_p}{8m^2_e}\left[(C_v+C_A)^2+(C_V-C_A)^2\left(1-\frac{T_p}{E_\nu}\right)^2\right. \\
\left.-(C^2_V-C^2_A)\frac{m_pT_p}{E^2_\nu}\right],
\end{split}
\end{equation}
where $m_p$ and $T_p$ are the mass and recoil kinetic energy of the proton respectively, $C_V=1/2-2\sin^2\Theta_W$ and $C_A=1.27/2$. The same expression holds for antineutrinos with $C_A=-C_A$. For a neutrino of energy $E_\nu$ the maximum proton kinetic energy $T^{max}_p$ is
\begin{equation}
T^{max}_p=\frac{2E^2_\nu}{m_p+2E_\nu}.
\end{equation}
 In a scintillator the light output from low energy protons is reduced relative to the light output for an electron depositing the same amount of energy. Taking into account this proton quenching the threshold on the proton kinetic energy in KamLAND is $T^{min}_p$=1.2 MeV \cite{kamland}. Therefore we integrate the differential cross section as
 \begin{equation}
\sigma=\int^{T^{max}_p}_{T^{min}_p} \frac{d\sigma_{\nu p}}{dT_p},
\end{equation}
and when replaced in Eq.\ref{ratecount}, $\sigma$ is integrated with respect to $E_\nu$ with
\begin{equation}
E_{th}\approx\left[\frac{m_pT^{min}_p}{2}\right]^{1/2}.
\end{equation}

We obtain the total counts by multiplying the rate count $R$ by the duration of the signal.  This time is determined by the amount of the total binding energy, $E_B$, that can be emitted by neutrinos. We note this quantity as $E^\nu_B$. McLaughlin and Surman \cite{McLaughlin07} estimated that  from the total binding energy at $r_s$, $E_B=9(M/M_\odot)\times 10^{53}$ ergs ($M$ being the NS mass), 20$\%$  is released in the form of neutrinos. This means $E^\nu_B\sim0.1Mc^2$. Other estimates of accretion onto BHs result in a budget of $GM/r_{ms}\sim0.1c^2=10^{20}$ergs/g, with $r_{ms}$ the radius of the marginally stable orbit \cite{Lee-Ramirez}, or, if a relativistic disk accretion onto a Schwarzschild BH is used $5.7\%$ of the rest mass energy \cite{Shapiro}. 

Roughly speaking, neutrinos would be emitted during an interval of time $dt=E^\nu_B/L$, with $L$ the total neutrino luminosity. However, due to the strong gravitational field caused by the BH, observers on Earth will detect a longer signal lasting an interval $dt^*$,
\begin{equation}
dt^*=\frac{dt}{|g_{00}|^{1/2}}.
\end{equation}
To estimate an overall  $|g_{00}|^{1/2}$, we use our results for the observed and emitted energy averages and the relation  between emitted and observed energies, Eq. \ref{shiftenergy}.  Using the estimates for $E^\nu_B$, we find that observers on Earth will detect a signal lasting for an interval of time $dt^*=0.15-0.07$ s. The estimate of the neutrino signal for a supernova is ~10 s and for AD-BH from NS-NS mergers $\sim 1$ s \cite{Dessart09}. Therefore the AD will emit a signal approximately 100 times shorter but, according to the results in table \ref{energytable}, more luminous. These fact offers the possibility of distinguishing the AD-BH spectrum from that of PNS. 

Table \ref{tablecounts} shows our results for the 

BH-NS neutrino  counts registered at the several detectors discussed in this section,  with an estimated signal time $dt^*= 0.15$s. Unless indicated we have added results over all neutrino flavors, and/or over the final products. For example reactions on ICARUS correspond to the sum of counts for absorption of $\nu_e$ plus ${\bar\nu_e}$. For comparison purposes we also show the current estimates for a PNS. From table \ref{tablecounts}, we can see that, in most of the cases, an AD-BH located a 10 kpc will generate more counts that a PNS at the same distance. This is general for all the detectors considered here.  In particular SK will record 1000 more events from 

 a BH-NS merger.

\begin{table}
\begin{center}
\caption{Neutrino counts from a 
BH-NS merger as registered at several facilities. Lines between rows separate detectors according their principle of detection. The size of the detectors are indicated in parenthesis. We include the PNS count estimates for some detectors also indicated in parenthesis.}

\begin{tabular}{llllllll}
\hline
&$ \bar{\nu}_e +p\rightarrow n+e^+$& $\nu+e\rightarrow\nu+e$\\
 SK(32 kton)& 9100&390 \\
UNO(580 kton)&165000&7100\\
Hyper-K(1Mton)&284000&12280\\
Amanda(680 OM)& 74000 &  2800 &  &\\
IceCube(4800 OM)& 522500 &  20200 &  &\\
PNS(SK)&8300&320\\
\hline
&$ \nu+p\rightarrow\nu+p$\\
 KamLAND (1 kton)& 470\\
 PNS&273  \\
 \hline
&$\nu_e+^{208}$Pb&$ \nu+^{208}$Pb&total\\
&$\rightarrow^{207(6)}$Bi+$e$&$\rightarrow^{207(6)}$Pb\\
  HALO (80 ton)&24&23&47\\
       PNS&&&43\\   

  \hline
  &$\nu_e(\bar{\nu}_e)+^{40}$Ar & $\nu+e\rightarrow\nu+e$\\
  &$\rightarrow e(e^+)$ &&\\
  ICARUS (3 kton)&331&30\\
   LANNDD(70 kton)&7700& 700\\
     PNS(ICARUS)&203&41\\
\hline
\end{tabular} 
\label{tablecounts}
\end{center}
\end{table}

Using the neutrino surfaces of the steady state disk model we estimate the counts obtained in SK. We predict an event rate of $8000$/s and $31000$/s for $a=0$ and $a=0.95$ respectively. The higher rates are a result of the higher temperatures of the rotating disk. To estimate the total counts at a given detector we need to estimate the binding energy available for neutrino emission. As in the previous case, to calculate the duration of the signal, we assume that the total mass dragged to the BH corresponds to 1.6$M_\odot$ and then divide by the total neutrino luminosity. This give us a total of 5400 counts in SK  for $a=0$ and 4000 for $a=0.95$. The total number of counts in the rotating disk are less because such disk is more luminous and therefore the available binding energy radiated as neutrinos would be consumed faster than in the case of a non-rotating disk. The extent to which steady state and dynamical models can converge on similar structure is not yet clear. We continue our calculations with the multidimensional model. As the disks that occur in compact object mergers become better understood, the methodology presented here can be applied to these new disk models as well.

Re-scaling to different distance we can use these results to determine how far we can see neutrinos from NS-BH accretion disks. For example, to detect neutrino from a supernova in AMANDA a vast amount of counts is needed, so the signal stands out from the background noise \cite{amanda}. If we speculate that for being able to distinguish the signal from 
 BH-NS merger, then it is necessary at least the same amount of counts as from supernova at 10 kpc, then AMANDA could reach AD-BH as far as 30 kpc. This estimate assumes a window interval of 10 sec; however the 
 BH-NS
signal lasts around 0.1 sec, therefore this 

 is a conservative estimate.
A more optimistic figure is obtained if we consider a large scale detector like UNO, which would detect at least one count from AD-BHs located as far as 4 Mpc. Compare this result with Advance LIGO, whose reach for NS-NS is 300 Mpc and for NS-BH is 650 Mpc \cite{ALIGO}. In order to see a NS-BH merger in the same distance range, but in neutrinos, requires a Gigaton scale detector.

\section{Neutrino Mixing}
\label{Neutrino Mixing}
In their way from the accretion disk toward a detector on Earth neutrinos will go through flavor transformation. At present there is some uncertainty in the flavor transformation, in part due to unknown neutrino parameters such as the hierarchy and the third mixing angle, and in part due to the lack of a complete calculation of neutrino flavor transformation for neutrinos leaving accretion disks.  However, since most detectors measure flavor dependent signals, the neutrino oscillations will have some effect on the number of counts. In this section we briefly review the possibilities for flavor transformation, and then take a few scenarios to demonstrate the range of possibilities. We follow the procedure of Kneller {\it et al.} \cite{Kneller08}.

Neutrinos have mass and therefore the weak states $e, \mu, \tau$ can be described as linear combinations of the mass states $m_1, m_2, m_3$. In the presence of matter, the Hamiltonian describing the evolution of the neutrino states is neither diagonal in the flavor basis nor in the mass basis. The coefficients describing the linear combination in any basis will oscillate, and their behavior will depend on the medium through which the neutrinos propagate, their energy, the differences between the squares of the masses $\delta m^2_{ij}=m^2_{i}-m^2_{j}$ and the mixing angles connecting the flavor and mass basis. The matrix $U$ relating the flavor and mass states can be written as

\begin{center}
\begin{widetext}
\begin{equation}
U=\begin{pmatrix}
c_{12}c_{13}&s_{12}c_{13}& s_{13}e^{-i\delta}\\
-s_{12}c_{23}-c_{12}s_{12}s_{23}e^{i\delta}&c_{12}c_{23}-s_{12}s_{13}s_{23}e^{i\delta}&c_{13}s_{23}\\
s_{12}s_{23}-c_{12}s_{13}c_{23}e^{i\delta}&-c_{12}s_{23}-s_{12}s_{13}c_{23}e^{i\delta}&c_{13}c_{23}
\end{pmatrix}.
\end{equation}
\end{widetext}
\end{center}

The coefficients of $U$ depend on the mixing angles $\theta_{12},\theta_{13}$, and $\theta_{23}$, and the CP- violating phase $\delta$. In $U$, $c_{ij}=\cos\theta_{ij}$ and $s_{ij}=\sin\theta_{ij}$. So far there is not a way to discriminate the  neutrino mass ordering of the neutrinos. The $m^2_1<m^2_2<m^2_3$ relation is known as the normal hierarchy (NH) while $m^2_3<m^2_1<m^2_2$ is refered to as the inverted hierarchy (IH). We consider oscillation scenarios for both  NH and IH. 

For a supernova density profile there is one matter resonance at low densities, usually called the $L$ resonance, and another matter resonance for high densities, the $H$ resonance. The $L$ resonance mixes the mass states $\nu_1$ and $\nu_2$ involving the mass difference $\delta m^2_{12}$, which can be determined by solar neutrino experiments $\delta m^2_{12} \backsim \delta m^2_\odot$ and $\theta_{12}\backsim \theta_\odot$. Because of the uncertainty in the mass hierarchy the $H$ resonance can involve different mass states. In the normal hierarchy the states mixed are $\nu_2$ and $\nu_3$, and involves the mass splitting $\delta m^2_{23}$ which is unknown. In the inverted hierarchy the mixed states are $\bar \nu_1$ and $\bar \nu_2$ involving the mass splitting $\delta m^2_{13}$. For both hierarchies the mixing angle is the small  $\theta_{13}$. The experimental limit to 

date is $\sin^22\theta_{13}<0.19$ \cite{PDG}. The probability for a neutrino going from one matter state to the other in $L$ resonance is denoted here as $P_L(E_\nu)$, and in the $H$ resonance as $P_H(E_\nu)$ and $\bar P_H(E_{\bar \nu})$. Antineutrinos do not cross from one state to another in the $L$ resonance. If the density changes slowly the neutrinos can propagate adiabatically. Therefore a neutrino produced in a 

matter
state will remain in the same 

matter
state as propagates through the medium. This means the crossing probability is close to zero and the dominant flavor state after passing through the resonance would have changed. The resonance is then ``adiabatic". A ``non-adiabatic" resonance corresponds to a crossing probability closer to one.

 Neutrino self-interactions can also affect the probability of detecting one flavor neutrino or another, and this is important in the region relatively close to the emission point where these self-interactions are large.  For a review see \cite{Duan:2009cd}.  As described in the review, depending on the density profile, the luminosities of the neutrinos, and the neutrino parameters, e.g. the hierarchy, these self-interactions can produce a range of new behaviors.  These include a spectral swap, a spectral split, and no change at all.  We follow the expressions in Ref. \cite{Kneller08}, and denote the survival probablities from this region where self-interactions can dominate as  $P_{SI}(E_\nu)$ and $\bar P_{SI}(E_{\bar\nu})$.

While ideally one should follow the evolution of the neutrino wave functions fully as they travel out of the accretion disk, as long as the relevant, H, L and SI regions are well separated we can approximate the  
survival neutrino and antineutrino probabilities, $p$, $\bar p$, as follows.

In the normal hierarchy,
\begin{widetext}
\begin{eqnarray}
p&=&\left[|U_{e1}|^2P_L+|U_{e2}|^2(1-P_L)\right] \left[P_H(1-P_{SI})+(1-P_H)P_{SI}\right]+|U_{e3}|^2\left[P_HP_{SI}+(1-P_H)(1-P_{SI})\right]\nonumber\\
{\bar p}&=&|U_{e1}|^2\left[\bar P_H\bar P_{SI}+(1-\bar P_H)(1-\bar P_{SI})\right]+|U_{e3}|^2\left[\bar P_H(1-\bar P_{SI})+(1-\bar P_H)\bar P_{SI}\right],
\end{eqnarray}
and in the inverted hierarchy,
\begin{eqnarray}
p&=&\left[|U_{e1}|^2P_L+|U_{e2}|^2(1-P_L)\right] \left[P_HP_{SI}+(1-P_H)(1-P_{SI})\right]+|U_{e3}|^2\left[P_H(1-P_{SI})+(1-P_H)(1-P_{SI})\right]\nonumber\\
\bar p&=&|U_{e1}|^2 \left[\bar P_H(1-\bar P_{SI})+(1-\bar P_H)\bar P_{SI}\right]+|U_{e3}|^2\left[\bar P_H\bar P_{SI}+(1-\bar P_H)(1-\bar P_{SI})\right].
\end{eqnarray}
\end{widetext}

The observed fluxes, after flavor oscillations, are expressed in terms of the survival probabilities as
\begin{eqnarray}
\phi_{\nu_e}&=&p\phi^0_{\nu_e}+(1-p)\phi^0_{\nu_x},\\
\phi_{\bar \nu_e}&=&\bar p\phi^0_{\bar {\nu}_e}+(1-\bar p)\phi^0_{\bar{\nu}_x},\\
\phi_{\nu_x}&=&\frac{1}{2}\left[(1-p)\phi^0_{\nu_e}+(1+p)\phi^0_{\nu_x}\right],\\
\phi_{\bar \nu_x}&=&\frac{1}{2}\left[(1-\bar p)\phi^0_{\bar{\nu}_e}+(1+\bar p)\phi^0_{\bar{\nu}_x}\right],
\end{eqnarray}
where $\phi^0$ are the initially produced fluxes and $x$ is either of the heavier flavors, $x=\tau=\mu$. Similarly, using the initially emitted energies the results of flavor oscillations on the energies. We insert the corresponding fluxes, including oscillations, in Eq. \ref{ratecount} and calculate the counts at the detector.  We consider oscillations scenarios with both normal and inverted hierarchies. For simplicity we calculate the survival probabilities assuming completely adiabatic or non-adiabatic resonances, that is $P_L=0$ or $P_L=1$, and similarly for $P_H$ and $P_{SI}$. Combining all the values of $P_H$ and $P_L$ we have a total of 8 scenarios. Finally,  we use 
$\sin^2\theta_{12}=0.311$ and $\sin^2\theta_{13}=10^{-4}$ \cite{PDG} for the mixing angles needed to calculate the matrix coefficients $U_{ij}$.

From the different scenarios studied we chose the ones which give extreme values.  Therefore, we can have an idea of what could be the upper and lower limit of our estimates and how much oscillations can affect them. Several scenarios reproduce the same values. We pick two for our discussion, S1 and S2. S1 correspond to a completely adiabatic H resonance ($P_H=\bar P_H=0$) and non-adiabatic L resonance ($P_L=1$) in the NH. S2 considers IH, with $P_L=P_H=\bar P_H=0$. We have used  $\bar P_{SI} = 0$, and $P_{SI}=1$ for both NH and IH.  Table \ref{tablefluxosc} shows the results of neutrino oscillations for energies and number neutrinos emitted per unit time $f$. Table \ref{tablecountosc} shows the effects of neutrino mixing in the registered events for the same oscillation scenarios.
The energies of $\nu_e$ and $\bar\nu_e$ are larger while energies for $\nu_x$ and $\bar\nu_x$ decrease. As a consequence of the neutrino mixing the energies of all neutrino flavors get closer together and the values of the fluxes also become closer together.
Counts in most detectors are larger as a result of neutrino mixing, except in KamLAND where our results remain constant. The extra counts in detectors where initially an electron neutrino or antineutrino is scattered can be easily understood: there is an extra contribution to the flux from $(\bar\nu_x)\nu_x$ emitted at high temperatures from the neutrino surface, see table \ref{energytable}.
\begin{table}
\begin{center}
\caption{Effects of neutrino mixing in the number of neutrinos emitted per second $f$ and in energy. We have included the values before mixing for comparison (NOsc).}
\begin{tabular}{llllllll}
\hline
&NOsc& S1& S2\\
(MeV)\\
$ E_{\nu_e}$& 17.3&20&26\\
$E_{\bar{\nu}_e }$&23.4&24&26\\
$ E_{\nu_x}$&26&25&22\\
$E_{\bar{\nu}_x}$&26&25.6&25\\
\hline
($\times10^{57} \nu$/s)\\
$ f_{\nu_e}$& 6.0&6.4&7.1\\
$f_{\bar{\nu}_e }$&6.5&6.7&7.1\\
$f_{\nu_x}$&7.1&6.9&6.6\\
$f_{\bar{\nu}_x}$&7.1&7.0&6.8\\
 
\hline
\end{tabular} 
\label{tablefluxosc}
\end{center}
\end{table}

\begin{table}
\begin{center}
\caption{Effects of neutrino mixing in events registered at several facilities. We have included the values before mixing for comparison (NOsc). For ICARUS we only show counts due to the current charged channel (CC)}
\begin{tabular}{llllllll}
\hline
&NOsc& S1& S2\\
$ \bar{\nu}_e +p\rightarrow n+e^+$\\
SK&9100&9800&11460\\
AMANDA&74000&79970&93200\\
\hline
$\nu+e\rightarrow\nu+e$\\
SK&390&430&490\\
\hline
ICARUS (CC)&331&450&710\\
HALO &47&62&91\\
KamLAND&470&468&467\\ 
\hline
\end{tabular} 
\label{tablecountosc}
\end{center}
\end{table}

\section{Summary and Conclusions}
\label{Conclusions}
The physics of NS-BH mergers is very important for our understanding of GRBs and the production of rare isotopes. 

Using a state-of-the-art model for a NS-BH merger we have calculated neutrino surfaces, fluxes luminosities and energy averages for the neutrinos emitted from the emerging AD. Neutrino surfaces as seen from the $z$ axis have an irregular shape. Due to large cross sections the surface for $\nu_e$ extends higher above the disk plane compared to the other neutrino flavor surfaces. We find that the redshifted neutrino energies are higher than in the case of a PNS. Our estimates are $E_{\nu_e} \approx 17.3 < E_{\bar{\nu_e}}  \approx 23.4<E_{\nu_x} \approx 26$ MeV, which include redshift due to the strong BH gravitational field. The AD-BH neutrino luminosity from the NS-BH merger is also larger than the one of a PNS. We find a total neutrino luminosity $\sim 10^{54}$ ergs/s; this is two orders of magnitude larger than the luminosity of a PNS. According to previous works this luminosity could be enough to trigger short GRB \cite{SetiawanBHmerger}.

These results are based on a snapshot of the disk evolution. We have calculated the same quantities for a steady state model and found that the average energies are lower compared to the hydrodynamical model. This is due to the contribution of low temperature neutrinos emitted at large radial distances, $\rho$, from the disk to the spectra in the steady state cases. We explored the influence of the spin parameter in the steady state disk using two different values.  We have found that higher spin parameters lead to higher neutrino energies and luminosities. As a consequence the neutrino counts per second are larger for the rapid rotating disk. However, if the fraction of binding energy released in neutrinos were the same for both, rotating and non-rotating disks, the signal in the rotating disk will be shorter resulting in less total counts. The order of magnitude of the luminosity for the steady state disk with $a$=0.95 is the same as for the 3D model and as it has been discussed could be a good candidate for triggering GRB \cite{SetiawanBHmerger}.

As part of our estimates, we have explore the consequences of neutrino mixing on the neutrino energies and events registered. For this purpose, we consider a total of 8 oscillation scenarios, which include the normal and inverted hierarchy.  This demonstrates the range of possible fluxes and energies that can be expected for different flavors. These estimates of the luminosity and energies of neutrinos from the disks of BH-NS mergers have a number of applications.
  
Based in our results for the neutrino surface temperatures, we estimate the number of events that would be registered in several detectors, both proposed and in operation if a  BH-NS were to occur in the galaxy.  While such events are expected to be rare in comparison with the standard core collapse supernova, it is still important to understand the potential signal.  Our estimates can be used to rule out this object as the origin of a future neutrino signal, as well as to guide investigations of neutrino detectors for the distant future. For these estimates we have included corrections due to the BH gravitational field such as energy redshift and bending of the neutrino trajectories. In general we find that there will be more counts coming from an AD-BH that is formed in a neutron star - black hole merger than counts registered from a PNS. For example in SK we will register 1000 more counts if a NS-BH merger happens than if a supernova event occurs. However, the neutrino signal from AD-BH would last only 0.15 s compared to the 10 s estimate for a PNS. We have then a signal that is 100 times more luminous  and 100 times shorter than a PNS.  We warn that our estimated counts are sensitive to the time the signal lasts. And our estimate for the duration depends on the efficiency of converting gravitational energy into neutrino energy. Therefore, the total counts could change as much as a factor of two. Nevertheless, our results for neutrino energies, luminosity and fluxes are independent of this uncertainty. The amount of AD-BH neutrinos that could be detected in large scale facilities and the possible distinction of this signal from a PNS opens the question: Is there a way to distinguish the neutrinos from a BH-NS merger and a supernova? The timing of the signal and the energies of the neutrinos will be different, so this is an important clue.  Therefore, it is not likely that we will mistake neutrinos from a merger event as those from a core collapse supernova.  How could we guarantee a detection of BH-NS neutrinos? 
So far with a large scale facility like UNO we could detect neutrinos from AD-BH as far 4 Mpc, which covers galaxies in the local group such as Andromeda. To detect a BH-NS merger in neutrinos and compete with a gravitational wave detector such as Advance LIGO would take a three orders of magnitude larger detector.

\section{Acknowledgments}
O. L. C. thanks David Brown for interesting discussions and John Blondin for help with the 3D visualization software. This work was supported in part by DOE contract DE-FG02-02ER41216.

\vfill\eject

\end{document}